\documentclass{article}

\usepackage{arxiv}

\usepackage[utf8]{inputenc} 
\usepackage[T1]{fontenc}    
\usepackage{hyperref}       
\usepackage{url}            
\usepackage{booktabs}       
\usepackage{amsfonts}       
\usepackage{nicefrac}       
\usepackage{microtype}      
\usepackage{lipsum}
\usepackage{graphicx}
\usepackage{subfigure}
\graphicspath{ {./images/} }
\usepackage{amsmath}
\usepackage{multicol}
\usepackage{multirow}

\title{A novel state connection strategy for quantum computing to represent and compress digital images}

\author{
Md Ershadul Haque \\
  School of Computing, Mathematics and Engineering\\
  Charles Sturt University\\
  Bathurst, NSW 2795 \\
  \texttt{mhaque@csu.edu.au} \\
   \And
Manoranjan Paul \\
  School of Computing, Mathematics and Engineering\\
  Charles Sturt University\\
  Bathurst, NSW 2795 \\
  \texttt{mpaul@csu.edu.au} \\
  \And
 Anwaar Ulhaq \\
  School of Computing, Mathematics and Engineering\\
  Charles Sturt University\\
  Bathurst, NSW 2795 \\
  \texttt{aulhaq@csu.edu.au} \\
    \And
Tanmoy Debnath \\
  School of Computing, Mathematics and Engineering\\
  Charles Sturt University\\
  Bathurst, NSW 2795 \\
  \texttt{tdebnath@csu.edu.au} \\
}
\begin{document}
\maketitle
\begin{abstract}
Quantum image processing draws a lot of attention due to faster data computation and storage compared to classical data processing systems. Converting classical image data
into the quantum domain and state label preparation complexity is still a challenging issue. The existing techniques normally connect the pixel values and the state position directly. Recently, the EFRQI (efficient flexible representation of the quantum image) approach uses an auxiliary qubit that connects the pixel-representing qubits to the state position qubits via Toffoli gates to reduce state connection. Due to
the twice use of Toffoli gates for each pixel connection still
it requires a significant number of bits to connect each pixel
value. In this paper, we propose a new SCMFRQI (state connection
modification FRQI) approach for further reducing the
required bits by modifying the state connection using a reset
gate rather than repeating the use of the same Toffoli gate
connection as a reset gate. Moreover, unlike other existing
methods, we compress images using block-level for further reduction
of required qubits. The experimental results confirm
that the proposed method outperforms the existing methods
in terms of both image representation and compression points
of view.  
\end{abstract}
\section{Introduction}
\label{sec:intro}

In quantum information processing (QIP)  quantum mechanics mainly deals with quantum properties \cite{b1}. Entanglement, parallelism, and superposition are the main properties in quantum mechanics that provide faster computation \cite{b2, b3,b4}. Storage, faster computation, and computational complexity are the main limitations of classical computers \cite{bb5}. Since the number of transistors, every two years will double in the past decade and after that has not increased significantly \cite{bb4}. Feynman et al. explored the first quantum computer and other ways to increase computing power \cite{bb5}.In \cite{b5,b6}, Shor and Grover, proposed an algorithm for factorial and database search in the quantum computer which showed faster computation than a classical computer \cite{b6}. 

In many applications, image processing plays core characteristics of image operation \cite{bb910,b7}. To convert pixel domain image into the quantum image, a number of approaches were already introduced \cite{sang2017novel,su2021improved,zhang2013neqr}. The main idea of the conversion is to represent the pixel value and the corresponding position using qubits and then make a connection using the operational gates between the pixel values, and their corresponding state position. Most of the existing models connect the pixel values representing qubits to the position values representing qubits directly\cite{b8}. For this reason, a huge number of gates is necessary to make the complete connection to present a real image. To address this issue, the EFRQI technique adds an auxiliary qubit (please see Fig. 1 for an example) to connect the pixel representing qubits to the state position qubits via Toffoli gates. Through this process, the EFRQI reduces the number of gates compared to the other methods, however, still, it requires a significant number of gates i.e. bits due to the twice use of Toffoli gates for each pixel connection \cite{b9}. To address the higher bit rate issue of the EFRQI, the proposed SCMFRQI approach uses only one reset gate which reduces the number of required gates significantly. The reset gate is used to nullify the effect of the present preparing pixel values on the next pixel value preparing operation. The rest of the article is organized as follows. The literature survey is discussed in Section \ref{L_R}; the proposed methodology is presented in section \ref{P_M}; the result and its discussion are given in section \ref{R_D}. The conclusion of this work is outlined in section \ref{CC}.

\section{Literature survey}\label{L_R}
Quantum computing (QC) is a combined application of quantum mechanics, computer science, and mathematics to increase the computational power\cite{b5}. Qubit lattice is the first quantum approach for image storing and retrieving \cite{b10}. Latorre et al. proposed the Real Ket approach to represent quantum image\cite{b12}. Inspired by the pixel-wise representation of the image in a classical computer, FRQI (Flexible Representation of Quantum Image) was developed in \cite{b13}.It encodes the color and position of the image using associated angle and kets. it was unable to represent pixel-wise grayscale complex operation since it used only one qubit. In addition, due to complexity of angle prepare to real image is really challenging. The Entanglement image representation was proposed in 2010 \cite{b14}. Based on JPEG(Joint Photography Expert Group) image, Jiang et al. proposed an efficient quantum compression method using the GQIR approach via DCT preparation method \cite{pm}. Laurel et al. proposed a quantum-based equivalence pixel image from bit pixel image \cite{b15}. The NEQR (Novel Quantum Representation of Color Digital Images), was proposed to represent the classical color image \cite{b16}. A NEQR method resolves the FRQI issues because it provides several qubits to represent the grayscale image. The limitation of NEQR is that its only able to represent the square image. To solve this problem INEQR approach was proposed \cite{b9}. In INEQR, how color and the big-size image are represented is still unclear. A GQIR (General Quantum Image Representation) uses a logarithmic scale to represent the rectangular image of arbitrary shape but it generates a lot of redundant bits \cite{b18}. Polar coordinate-based quantum knows as QUALPI\cite{b19}. 

\begin{figure}[htbp]
\centerline{\includegraphics[width=\linewidth]{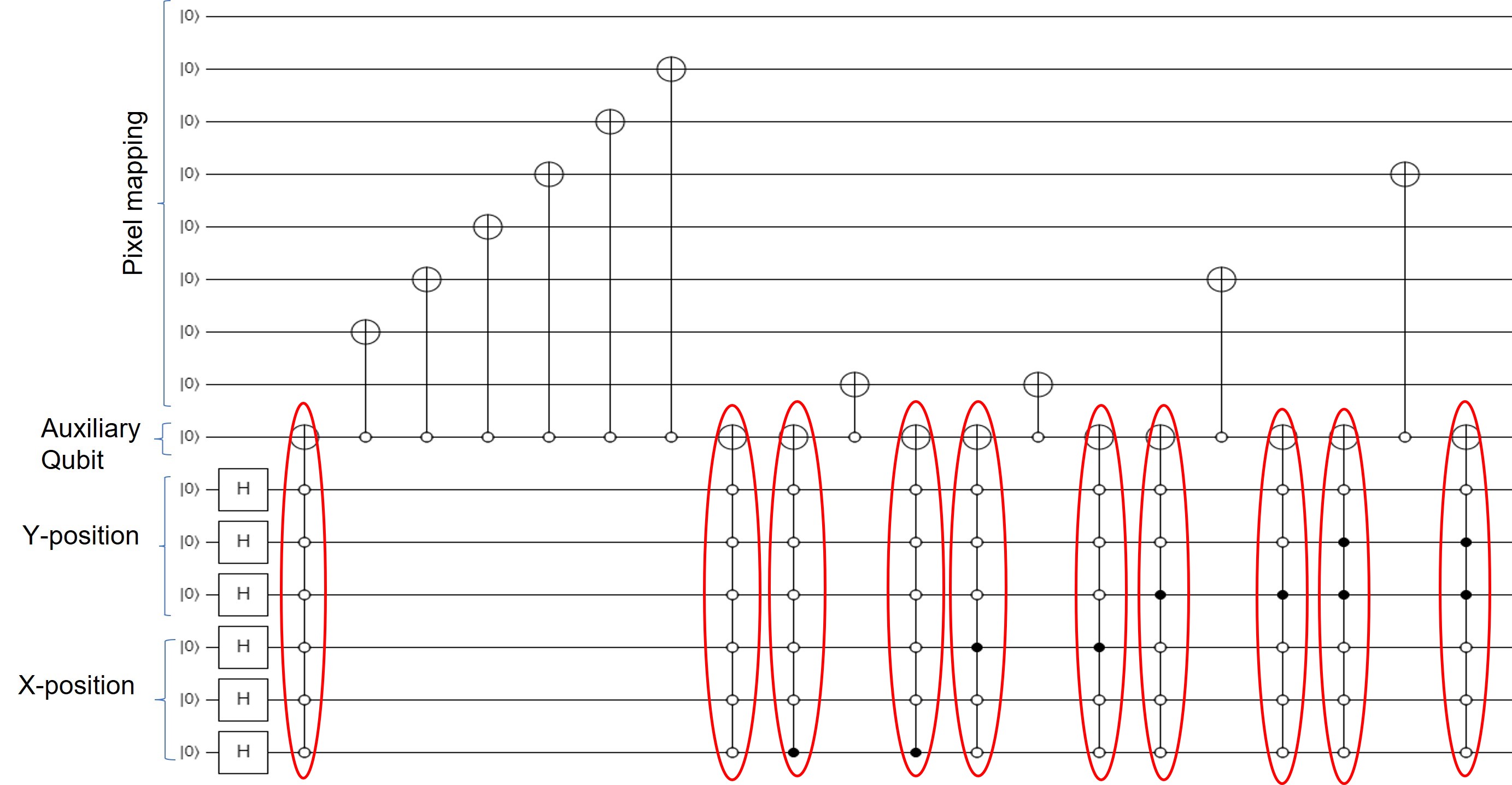}}
\caption{EFRQI circuit diagram for pixel values representation}
\label{fig_efrqi}
\end{figure}

In 2021, an EFRQI (Efficient Flexible Representations of Quantum Image) was proposed to minimize the state preparation bits \cite{b8}. Figure \ref{fig_efrqi} shows an example of EFRQI approach for pixel values of 125(X=0,Y=0), 1(X=1,Y=0), 1(X=4,Y=0), 4(X=0,Y=1), and 16(Y=3,X=0) respectively. For each pixel state preparation, an EFRQI approach uses the same Toffoli gate twice (shown in the red circle) for connecting the state values qubits to the pixel values qubits. To complete a real connection, those Toffoli gates generate a higher amount of bits which is the main drawback of state label preparation of the EFRQI  approach. 
\section{Proposed Approach}
\label{P_M}
\begin{figure}[htbp]
\centerline{\includegraphics[width=\linewidth]{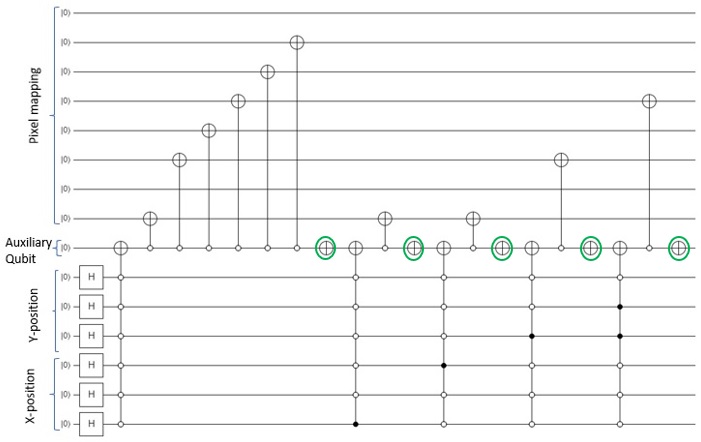}}
\caption{SCMFRQI circuit diagram for pixel values representation}
\label{fig_proposed_Chematic_diagram}
\end{figure}

Figure \ref{fig_proposed_Chematic_diagram} shows the circuit diagram of the proposed SCMFRQI scheme for representing pixels inside a quantum computer. Rather than using the same Toffoli gate, the SCMFRQI uses a single reset gate for each pixel connection(shown in the green circle). The Quirk simulation tools is used to map connecting circuit \cite{bs8}. 

A $2^a\times2^b$ image size is considered for representation and compression purposes.\\ 
In the SCMFRQI approach, the steps involved are given below: \\
Step 1: DCT and quantized. \\
Step 2: Prepare the quantized DCT coefficient. In SCMFRQI, it requires $q+2n+1$ qubits. Where $q$ is the number of required qubits to reprepixelsthe pixel or coefficient values and $n=log_2(S)$. One auxiliary qubit act as a bridging qubit to make the connection between pixel or coefficient preparation qubits and their corresponding position qubits. The initial state can be explained using the below equation\cite{pm}. 
 \begin{equation}
     \Psi_0\rangle={\vert0\rangle}^{\otimes(q+2n+1)}
 \end{equation}
Then, $(q+1)$ identity gates and $2n$ Hadamard gates are  used. In this step, the whole quantum step can be expressed as follows: 
\begin{equation}
U=I^{\otimes{q+1}}\otimes H^{\otimes{2n}}
\end{equation}
$U$ transform $\Psi_0$ from initial state to intermediate state $\psi_1$.
\begin{equation}
\Psi_1=U(|\Psi_0\rangle)=I|0\rangle^{\otimes{q+1}}\otimes H^{\otimes{2n}}
\end{equation}
The final preparation step is done using $U_2$ quantum operator:
\begin{equation}
\Psi_2=U_2(|\Psi_1\rangle)=\frac{1}{2^n} \sum_{i=1}\sum^{j=1}\,|C_{YX}\rangle |YX\rangle
\end{equation}
where $|C_{YX}\rangle$ is corresponding coefficient value and $YX$ its coordinate position. The quantum transform operator is $U_2$ is given below: 
\begin{equation}
U_2=\prod_{X=0,....,2^n-1}\prod_{Y=0,....,2^n-1}\, U_{YX}
\end{equation}
The quantum sub-operator $U_{YX}$ is also given below: 
\begin{equation}
U_{YX}= \left(I\otimes \sum_{ij\neq YX} {|ji\rangle {\langle ji|}} \right)   +\sigma_{YX} \otimes |YX\rangle {\langle YX|}
\end{equation}
The $\sigma_{YX}$ is given below:
\begin{equation}
    \sigma_{YX} =\otimes^{q-1}_{i=0}{\sigma^i_{YX}}
\end{equation}
The function of $\sigma^i_{YX}$ is setting the value of $i_{th}$ qubit of (YX)'s quantized DCT coefficient. \\
 Step 3: Store the quantized coefficient after performing $8X8$ block DCT.\\
 Step 4: Inverse quantization. \\
 Step 5: Inverse DCT. \\
 Step 6: Compute bit rate and PSNR. \\
 The connection of the Toffoli gate is given below.  
 \begin{equation}
          T_{bit}= (log_2(S_X)+log_2(S_Y)+1+1)\otimes{N_{tcn}}
 \end{equation}
 The required bit rate (BR) is calculated using the following equation:
 \begin{equation}
     {BR}=q_{ones}+S_{bit}+T_{bit}+A_{bit}+B_e
 \end{equation}
Where $q_{ones}$ is the total number of ones presented in the transfer coefficient or pixel values. A $S_{bit}$ is the sign bit that represents the sign of the non-zero pixel or coefficient values. A $N_{tcn}$ is the total number of non-zero coefficient or pixel elements. A $S_X$ and $S_Y$ represent the X and Y-position of nonzero pixel or coefficient and mapped by $4\times4$ quantum block-wise qubits system. An $A_{bit}$ is the required number of bits that come from the auxiliary qubits. Since the SCMFRQI approach uses $4\times4$ quantum block to locate the position values from $16\times16$ array image consequently, that is why $B_e$ is introduced to locate each block error. 
\section{Result and discussion}\label{R_D}
In this section, the experimental results are analyzed  for Deer$(1024\times1024)$, baboons$(512\times512)$, scenery $(512\times512)$ and peppers$(512\times512)$ images \cite{bb17,bb18} for  verification purpose of proposed approach. Four experiments have been conducted to demonstrate the proposed method's performance. The first experiment is taken via a direct approach, where the rest others are examined via DCT preparation approach.    

\subsection{Scenery I-result analysis of the direct approach}
\begin{figure}
\centering
    {
        \includegraphics[width=\textwidth]{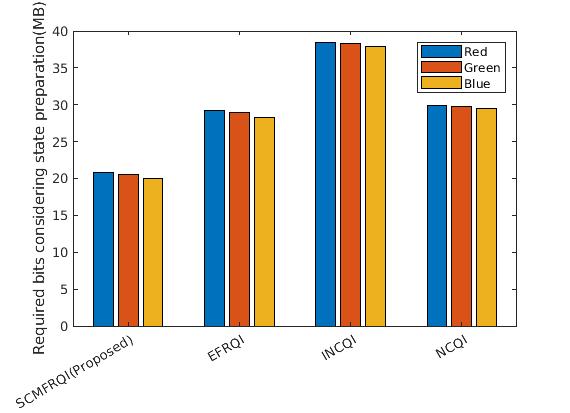}
        \label{color_channel_direct_Deer}
    }
    \caption{Proposed SCMFRQI scheme bit rate comparison}
    \label{B_PSNR_Four_image}
\end{figure}

Figure \ref{B_PSNR_Four_image} depicts that the required bit rates for the SCMFRQI scheme for R, G, and B channels of deer are approximately 22MB, 20MB, and 18MB respectively. On the other hand, the INCQI requires nearly 38MB, 37MB, and 36MB respectively. Besides, the NCQI requires a similar amount of bit rate to the EFRQI. 

For the baboon's image, Table \ref{Red_Green_Blue_direct} exhibits that the bit rate for R, G, and B channels is about 5.09MB, 5.02MB and 4.97MB  in the case of the SCMFRQI approach is less compared to EFRQI R(7.19MB), G(7.11MB),  B(7.06MB);  NCQI R(6.63MB), G(6.59MB), B(6.55MB);  INCQI R(8.72MB), G (8.69MB),B(8.64MB) respectively. For the scenery image, Table \ref{Red_Green_Blue_direct} shows that the required bit rate in terms of the SCMFRQI approach is R(4.97MB), G(4.99MB), B(4.93MB). The second highest required whereas of bit rates is found in terms of the EFRQI approach. On the other hand, NCQI requires on average 6.5MB bit rate for all kinds of channels. The blue channel comparison result from Table \ref{Red_Green_Blue_direct} shows that the SCMFRQI approach exhibits better results compared to others. 
\begin{table}[]
\centering
\caption{Bit rate for Baboons, Scenery and Peppers images}
\begin{tabular}{|cccc|c|c|}
\hline
\multicolumn{4}{|c|}{Bit rate(MB)}                                                                 & \multirow{2}{*}{\begin{tabular}[c]{@{}c@{}}Image\\  name\end{tabular}} & \multirow{2}{*}{Ch.} \\ \cline{1-4}
\multicolumn{1}{|c|}{SCMFRQI} & \multicolumn{1}{c|}{NCQI}   & \multicolumn{1}{c|}{INCQI} & EFRQI &                                                                        &                      \\ \hline
\multicolumn{1}{|c|}{5.09}  & \multicolumn{1}{c|}{6.63} & \multicolumn{1}{c|}{8.72} & 7.19 & \multirow{3}{*}{Baboons}                                               & Red                  \\ \cline{1-4} \cline{6-6} 
\multicolumn{1}{|c|}{5.02}  & \multicolumn{1}{c|}{6.59} & \multicolumn{1}{c|}{8.69} & 7.11 &                                                                        & Green                \\ \cline{1-4} \cline{6-6} 
\multicolumn{1}{|c|}{4.97}  & \multicolumn{1}{c|}{6.55} & \multicolumn{1}{c|}{8.64} & 7.06 &                                                                        & Blue                 \\ \hline
\multicolumn{1}{|c|}{4.97}  & \multicolumn{1}{c|}{6.54} & \multicolumn{1}{c|}{8.63} & 7.20 & \multirow{3}{*}{Scenery}                                               & Red                  \\ \cline{1-4} \cline{6-6} 
\multicolumn{1}{|c|}{4.99}  & \multicolumn{1}{c|}{6.55} & \multicolumn{1}{c|}{8.63} & 7.07 &                                                                        & Green                \\ \cline{1-4} \cline{6-6} 
\multicolumn{1}{|c|}{4.93}  & \multicolumn{1}{c|}{6.50} & \multicolumn{1}{c|}{8.57} & 7.00 &                                                                        & Blue                 \\ \hline
\multicolumn{1}{|c|}{5.10}  & \multicolumn{1}{c|}{6.63} & \multicolumn{1}{c|}{8.73} & 7.19 & \multirow{3}{*}{Peppers}                                               & Red                  \\ \cline{1-4} \cline{6-6} 
\multicolumn{1}{|c|}{4.76}  & \multicolumn{1}{c|}{6.32} & \multicolumn{1}{c|}{8.35} & 6.79 &                                                                        & Green                \\ \cline{1-4} \cline{6-6} 
\multicolumn{1}{|c|}{4.54}  & \multicolumn{1}{c|}{6.19} & \multicolumn{1}{c|}{8.21} & 6.55 &                                                                       & Blue                 \\ \hline
\end{tabular}
\label{Red_Green_Blue_direct}
\end{table}

\subsection{Scenery II-result analysis for the Red channel}
\begin{figure}
\centering
    \subfigure [Deer image]
    {
        \includegraphics[width=.45\textwidth]{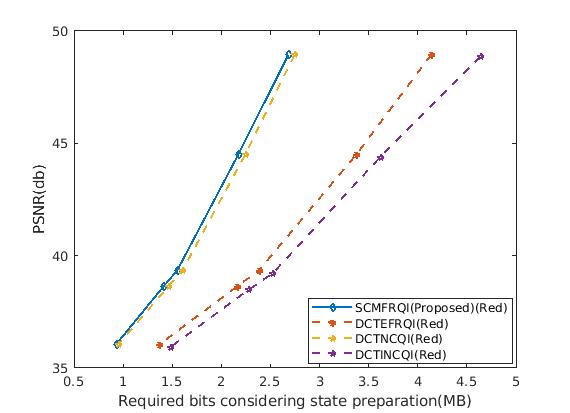}
        \label{red_channel_Deer}
    }
    \subfigure[Baboons image]
    {
        \includegraphics[width=0.45\textwidth]{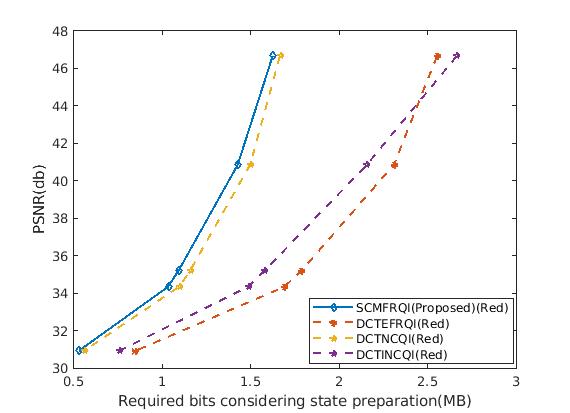}
        \label{red__channel_baboonsr}
    }
    \subfigure[Scenery image]
    {
        \includegraphics[width=0.45\textwidth ]{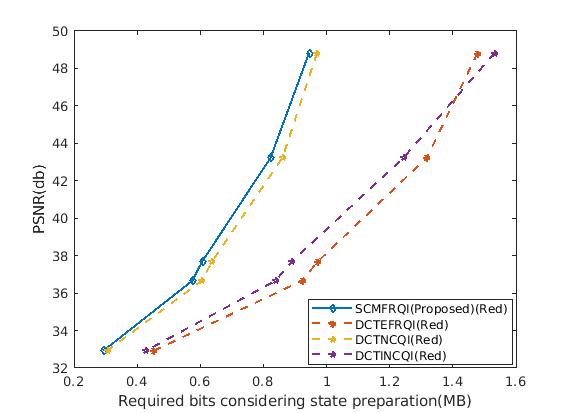}
        \label{red__channel_scenery}
    }
    \subfigure[Peppers image]
    {
        \includegraphics[width=0.45\textwidth ]{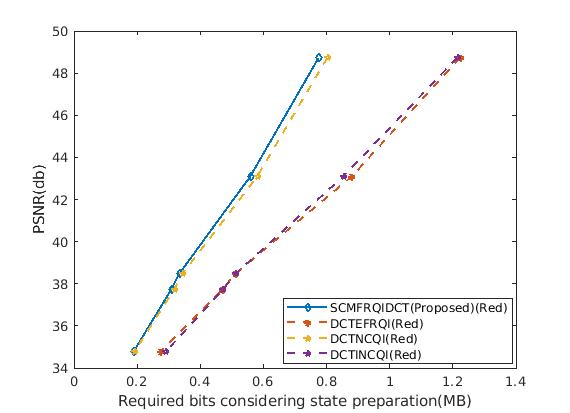}
        \label{red_channel_peppers}
    }
    \caption{Proposed scheme bit rate versus PSNR  for Red channel}
    \label{red_PSNR_comparison_color_image}
\end{figure}

Figure \ref{red_PSNR_comparison_color_image} shows the comparison result of SCMFRQI approach for the Red channel. Comparison results in \ref{red_channel_Deer} show that the proposed SCMFRQI requires lower bit rate compared to DCTINCQI, DCTNCQI, and DCTEFRQI respectively using  Q=8, 16, 32, 64, and 70 quantization factors. In the meantime, DCTNCQI requires higher bit rates compared to SCMFRQI where both  DCTEFRQI and DCTINCQI require more. Figure \ref{red__channel_baboonsr} exhibits the computational result which require less bit rate than DCTEFRQI, DCTINCQI, and DCTEFRQI approach's. The comparison results in Figure \ref{red__channel_scenery} depict that, the proposed SCMFRQI approach outlines the better result compared to DCTEFRQI, DCTINCQI, and DCTNCQI approaches while DCTNCQI shows the closest. In the meantime, DCTINCQI and DCTEFRQI require more bit rates versus PSNR values that are situated far away from the SCMFRQI approach. Figure \ref{red_channel_peppers} computation result of the proposed scheme for the Red channel of peppers scheme and compared to the existing considered methods. 

\subsection{Scenery III-result analysis for the Green channel}
Table \ref{Green_channel_SCMFRQI_DCT} shows the comparison results for the Green channel of Deer image, baboons, scenery, and peppers images for 8 quantization factors. For Deer image, comparison results show that the proposed scheme SCMFRQI is more capable to save bit rate compared to others. In terms of PSNR, its shows that all the methods draw a similar amount of PSNR values. Moreover, DCTINCQI and DCTEFRQI require a higher bit rate compared to the SCMFRQI scheme. For the baboons image, from Table \ref{Green_channel_SCMFRQI_DCT}, it is concluded that the proposed SCMFRQI scheme result shows more efficiency compared to all others considered approaches. On the other hand, in the case of scenery and peppers images, the SCMFRQI approach draws a better result compared to other approaches.

\begin{table}[]
\caption{Bit rate of Green channel via preparation approach}
\centering
\begin{tabular}{|c|cccc|c|}
\hline
\multirow{2}{*}{\begin{tabular}[c]{@{}c@{}}PSNR\\(db)\end{tabular}} & \multicolumn{4}{c|}{Bit rate (MB)}                                                                        & \multirow{2}{*}{Image} \\ \cline{2-5}
                          & \multicolumn{1}{c|}{SCMFRQI} & \multicolumn{1}{c|}{NCQI} & \multicolumn{1}{c|}{INCQI} & EFRQI &                             \\ \hline
48.99                   & \multicolumn{1}{c|}{2.59}  & \multicolumn{1}{c|}{2.65}   & \multicolumn{1}{c|}{4.51}    & 4.11    & \multirow{2}{*}{Deer}       \\ \cline{1-5}
39.29                   & \multicolumn{1}{c|}{1.56}  & \multicolumn{1}{c|}{1.62}   & \multicolumn{1}{c|}{2.55}    & 2.41    &                             \\ \hline
47.06                   & \multicolumn{1}{c|}{1.62}  & \multicolumn{1}{c|}{2.10}   & \multicolumn{1}{c|}{1.66}    & 2.65    & \multirow{2}{*}{Baboons}    \\ \cline{1-5}
35.23                   & \multicolumn{1}{c|}{1.10}  & \multicolumn{1}{c|}{1.49}   & \multicolumn{1}{c|}{1.59}    & 1.80    &                             \\ \hline
48.99                   & \multicolumn{1}{c|}{0.94}  & \multicolumn{1}{c|}{1.22}   & \multicolumn{1}{c|}{1.52}    & 1.54    & \multirow{2}{*}{Scenery}    \\ \cline{1-5}
37.75                   & \multicolumn{1}{c|}{0.60}  & \multicolumn{1}{c|}{0.79}   & \multicolumn{1}{c|}{0.88}    & 1.01    &                             \\ \hline
49.54                   & \multicolumn{1}{c|}{0.77}  & \multicolumn{1}{c|}{1.00}   & \multicolumn{1}{c|}{1.21}    & 1.27    & \multirow{2}{*}{Peppers}    \\ \cline{1-5}
38.67                   & \multicolumn{1}{c|}{0.34}  & \multicolumn{1}{c|}{0.44}   & \multicolumn{1}{c|}{0.52}    & 0.57    &                             \\ \hline
\end{tabular}
\label{Green_channel_SCMFRQI_DCT}
\end{table}

\subsection{Scenery IV-result analysis for Blue channel}
Table \ref{Blue_prepation} contains the comparative result of the Blue channel for Deer, Baboons, scenery, and peppers images. For all kinds of images, the performance result shows that the proposed SCMFRQI approach can represent and compress the images efficiently compared to the existing models in terms of bit rate and PSNR.

\begin{table}[]

\centering
\caption{Bit rate of Blue channel via preparation approach}
\begin{tabular}{|c|cccc|c|}
\hline
\multicolumn{1}{|l|}{\multirow{2}{*}{\begin{tabular}[c]{@{}c@{}}PSNR\\(db)\end{tabular}}} & \multicolumn{4}{c|}{Bit rate(MB)}                                                                                     & \multirow{2}{*}{\begin{tabular}[c]{@{}c@{}}Image\\  name\end{tabular}} \\ \cline{2-5}
\multicolumn{1}{|l|}{}                           & \multicolumn{1}{c|}{SCMFRQI} & \multicolumn{1}{c|}{NCQI}   & \multicolumn{1}{c|}{INCQI}  & EFRQI                       &                                                                        \\ \hline
49.00                                         & \multicolumn{1}{c|}{2.66}  & \multicolumn{1}{l|}{3.42} & \multicolumn{1}{l|}{4.29} & \multicolumn{1}{c|}{4.35} & Deer                                                                   \\ \hline
46.53                                          & \multicolumn{1}{c|}{1.63}  & \multicolumn{1}{c|}{2.12} & \multicolumn{1}{c|}{2.68} & 2.67                      & Baboons                                                                \\ \hline
48.49                                          & \multicolumn{1}{c|}{0.94}  & \multicolumn{1}{c|}{1.22} & \multicolumn{1}{c|}{1.53} & 1.55                      & Scenery                                                                \\ \hline
48.68                                          & \multicolumn{1}{c|}{0.77}  & \multicolumn{1}{c|}{1.02} & \multicolumn{1}{c|}{1.20} & 1.29                      & Peppers                                                                \\ \hline
\end{tabular}
\label{Blue_prepation}
\end{table}


%
%
%
\section{Conclusion}\label{CC}
In this work, through the analysis of the literate survey, the limitations of the existing quantum models have been figured out. Then a novel, SCMEFRQI approach has been proposed for any size of image representation and compression to improve the EFRQI approach which is the latest and best existing model for state connection. it uses an auxiliary qubit and resets gate that modifies the state connection and stores and compresses the color image information of all the pixels or transfer coefficient values instead of probability. Another advantages is that, it also avoid the complex rotation operation. Besides, it is more capable to represent and compress the image compared with existing models.      
\section*{Acknowledgment}
The author declare that there is no conflict of interest.

\bibliographystyle{unsrt}  


\begin{thebibliography}{40}

\bibitem{b1} 
Rabia Amin Khan. 
\newblock An improved flexible representation of quantum images.
\newblock \em {Quantum Information Processing}, vol. 18, no. 7, pages 1-19, 2019.

\bibitem{b2} 
Jacobs IS and  Bean CP.
\newblock Fine particles, thin films and exchange anisotropy.
\newblock in \em {Magnetism},  pages. 271--350, 1963.

\bibitem{b3} 
Thaddeus D. Ladd, Fedor Jelezko, Raymond Laflamme, Yasunobu Nakamura, Christopher Monroe, and Jeremy Lloyd O’Brien.
\newblock Quantum computers.
\newblock \em{nature}, vol. 464, no. 7285, pages. 45-53, 2010.

\bibitem{b4}
Mandra  Salvatore, Gian Giacomo Guerreschi, and Alán Aspuru-Guzik.
\newblock Faster than classical quantum algorithm for dense formulas of exact satisfiability and occupation problems. 
\newblock \em{New Journal of Physics}, vol. 18, no. 7, pages. 073003, 2016.

\bibitem{bb5}  
Zhaobin Wang, Minzhe Xu, and Yaonan Zhang. 
\newblock Review of quantum image processing.
\newblock \em{Archives of Computational Methods in Engineering}, pages. 1-25, 2021.

\bibitem{bb4} 
Richard P. Feynman.
\newblock Simulating physics with computers.
\newblock In \em{Feynman and computation}, pp. 133-153. CRC Press, 2018.

\bibitem{b5} 
 Peter W. Shor.
 \newblock Algorithms for quantum computation: discrete logarithms and factoring. 
 \newblock In \em{Proceedings 35th annual symposium on foundations of computer science}, pp. 124-134. Ieee, 1994.

\bibitem{b6} 
Grover, Lov K. 
\newblock A fast quantum mechanical algorithm for database search.
\newblock In \em{Proceedings of the twenty-eighth annual ACM symposium on Theory of computing}, pp. 212-219. 1996.

\bibitem{bb910}
 Kalonia, Mohini, and Trasha Gupta 
 \newblock Review on applications of quantum image processing.
 \newblock \em{Proc. SPIE 11878, Thirteenth International Conference on Digital Image Processing (ICDIP 2021)}, 118781T (30 June 2021), https://doi.org/10.1117/12.2601025.
 
\bibitem{b7} 
Gonzalez RC and Woods RE. 
\newblock Digital image processing.
\newblock \em{2ndEd, Prentice Hall}. 

\bibitem{sang2017novel} 
Sang, Jianzhi and Wang, Shen and Li, Qiong.
\newblock A novel quantum representation of color digital images
\newblock \em{{Quantum Information Processing}}, vol. 16, no. 2, pp. 1-14, 2017.

\bibitem{su2021improved} 
Su, Jie and Guo, Xuchao and Liu, Chengqi and Lu, Shuhan and Li, Lin.
\newblock An improved novel quantum image representation and its experimental test on IBM quantum experience,
\newblock \em{Scientific Reports}, vol. 11, no. 1, pp. 1-13, 2021.

\bibitem{zhang2013neqr}
Zhang, Yi and Lu, Kai and Gao, Yinghui and Wang, Mo
\newblock NEQR: a novel enhanced quantum representation of digital images
\newblock \em{Quantum information processing}, vol. 12, no. 8, pp. 2833-2860, 2013.

\bibitem{b8} 
 Norhan Nasr, Ahmed Younes and Ashraf Elsayed.
 \newblock Efficient representations of digital images on quantum computers. 
 \newblock \em{Multimedia Tools and Applications}, vol. 80, no. 25, pages.34019-34034, 2021.

\bibitem{b9} 
Nan Jiang and Luo Wang. 
\newblock Quantum image scaling using nearest neighbor interpolation. 
\newblock \em{Quantum Information Processing},vol. 14, no. 5, pages. 1559-1571, 2015.

\bibitem{b10} Venegas-Andraca, Salvador E., and Sougato Bose. 
\newblock Storing, processing, and retrieving an image using quantum mechanics.
\newblock In \em{Quantum Information and Computation}, Vol. 5105, pages. 137-147,  SPIE, Augest, 2003.

\bibitem{b12}
Jose I. Latorre.
\newblock Image compression and entanglement.
\newblock \em{arXiv preprint quant-ph/0510031}.

\bibitem{b13}
Phuc Q. Le, Fangyan Dong, and Kaoru Hirota.
\newblock A flexible representation of quantum images for polynomial preparation, image compression, and processing operations.
\newblock \em{Quantum Information Processing}, vol. 10, no. 1, pages. 63-84, 2011.

\bibitem{b14}
Venegas-Andraca, Salvador E.  and Ball JL.
\newblock Processing images in entangled quantum system.
\newblock \em{Quant Inf Process}, vol. 9, no. 1, pp. 1–11, 2010.

\bibitem {pm}
 Nan Jiang, Xiaowei Lu, Hao Hu, Yijie Dang and Yongquan Ca.
 \newblock A novel quantum image compression method based on JPEG.
 \newblock \em{International Journal of Theoretical Physics}, vol. 57, 3no. pp. 611-636, 2018.

\bibitem {b15}
Laurel, Carlos Ortega, Shi-Hai Dong, and M. Cruz-Irisson.
\newblock Equivalence of a bit pixel image to a quantum pixel image.
\newblock \em{Communications in Theoretical Physics}, vol. 64, no. 5, pages. 501, 2015. 

\bibitem {b16}
Jianzhi Sang, Shen Wang, and Qiong Li.
\newblock A novel quantum representation of color digital images.
\newblock \em{Quantum Information Processing},vol. 16, no. 2, pages. 1-14.

\bibitem{b18}
Jiang, Nan, Jian Wang, and Yue Mu.
\newblock Quantum image scaling up based on nearest-neighbor interpolation with integer scaling ratio.
\newblock \em{Quant Inf Process}, vol. 14, no. 11, pages. 4001–4026.

\bibitem{b19}
Zhang, Yi, Kai Lu, Yinghui Gao, and Kai Xu.
\newblock A novel quantum representation for log-polar images.
\newblock \em{Quant Inf Process}, vol. 12, no.9, pages. 3103–3126. 

\bibitem {bs8}
Intro to Quantum Software Development.
\newblock The Quirk Tool.
\newblock https://algassert.com/quirk.

\bibitem {bb17} 
USC University of South California. 
\newblock The USC-SIPI Image Database.
\newblock  (2022).

\bibitem {bb18}
Li, Yi.
\newblock Object and concept recognition for content-based image retrieval.
\newblock University of Washington. 

\end{thebibliography}

\end{document}